\documentclass[11pt]{article}

\usepackage{fullpage}
\usepackage{amsmath}
\usepackage{epsfig}
\newsavebox{\fmbox}
\newsavebox{\algobox}

\newcommand\eps\varepsilon

\newtheorem{theorem}{Theorem}

\newcommand\drop[1]{}
\newcommand\req[1]{(\ref{#1})}

\newcommand{\E}{\textnormal{E}}

\newcommand{\Var}{\textnormal{Var}}

\newcommand{\CoV}{\textnormal{CoV}}

\newcommand{\OPT}[1]{\textnormal{OPT}_{{#1}}}
\newcommand{\SYS}[1]{\textnormal{SYS}_{{#1}}}
\newcommand{\THR}[1]{\textnormal{THR}_{{#1}}}

\newcommand{\PRI}[1]{\textnormal{PRI}_{{#1}}}
\newcommand{\Uwor}[1]{\textnormal{U-R}_{{#1}}}
\newcommand{\Pwr}[1]{\textnormal{P+R}_{{#1}}}

\newcommand{\gappr}{\begin{array}{c}\\[-1em] >\\[-.55em]\sim\end{array}}
\newcommand{\VS}{{V\Sigma}}
\newcommand{\SV}{{\Sigma}V}
\newcommand{\Sp}[1]{{\mathcal S_{{#1}}}}
\newcommand{\Sm}[1]{{\mathcal S_{{#1}:n}}}
\newcommand{\Wp}[1]{W_{{#1}}}
\newcommand{\Vm}[1]{V_{{#1}:n}}
\newcommand{\MSEm}[1]{MSE_{{#1}:n}}
\newcommand{\MSES}{{MSE\Sigma}}
\newcommand{\SMSE}{{\Sigma}MSE}

\title{On the variance of subset sum estimation}

\author{Mario Szegedy\footnote{Research supported by NSF.}\\
Rutgers University\\
{\tt szegedy@cs.rutgers.edu}
\and Mikkel Thorup\\
AT\&T Labs---Research\\
{\tt mthorup@research.att.com}
}

\begin{document}

\maketitle

\begin{abstract}
For high volume data streams and large data warehouses, sampling is
used for efficient approximate answers to aggregate queries over selected
subsets. Mathematically, we are dealing with a set
of weighted items and want to support queries to arbitrary subset
sums. With unit weights, we can compute subset sizes which
together with the previous sums provide the subset averages. The question
addressed here is which sampling scheme we should use to get
the most accurate subset sum estimates.

We present a simple theorem on the variance of subset sum estimation
and use it to prove variance optimality and near-optimality of subset
sum estimation with different known sampling schemes. This variance is
measured as the average over all subsets of any given size. By optimal
we mean there is no set of input weights for which any sampling scheme
can have a better average variance. Such powerful results can never be
established experimentally. The results of this paper are derived
mathematically. For example, we show that appropriately
weighted systematic sampling is simultaneously optimal for all subset
sizes.  More standard schemes such as uniform sampling and
probability-proportional-to-size sampling with replacement can be
arbitrarily bad.

Knowing the variance optimality of different sampling schemes can
help deciding which sampling scheme to apply in a given context.
\end{abstract}



\section{Introduction}
Sampling is at the heart of many DBMSs, Data Warehouses, and Data
Streaming Systems.
It is used both internally, for query optimization, enabling selectivity
estimation, and externally, for speeding up query evaluation, and for
selecting a representative subset of data for
visualization~\cite{OR95}.
Extensions to SQL to support sampling are present in
DB2 and SQLServer (the TABLESAMPLE keyword \cite{ibm}), Oracle (the SAMPLE
keyword \cite{oracle}), and can be simulated for other systems using syntax such as
ORDER BY RANDOM() LIMIT 1.
Users can also ensure sampling is used for query optimization, for example
in Oracle (using { \tt dynamic-sampling} \cite{oracle-ds}).

Mathematically, we are here dealing with a set of weighted items and
want to support queries to arbitrary subset sums. With unit weights,
we can compute subset sizes which together with the previous sums
provide the subset averages. The question addressed here is which
sampling scheme we should use to get the most accurate subset sum
estimates. More precisely, we study the variance of sampling based
subset sum estimation. We note that there has been sevaral previous
works in the data base community on sampling based subset sum
estimation (see, e.g., \cite{ADLT05,HHW97,JMR05}).

The formal set-up is as follows.  We are
dealing with a set of items $i\in [n]$ with positive weights $w_i$. 
Here $[n]=\{1,...,n\}$. A
subset $S\subseteq[n]$ of these are sampled, and each sampled item $i$
is given a weight estimate $\hat w_i$. Unsampled items $i\not\in S$
have a zero weight estimate $\hat w_i=0$.  We generally assume that
sampling procedures include such weight estimates. We are mostly
interested in unbiased estimation procedures such that 
\begin{equation}\label{eq:single}
\E[\hat w_i]=w_i\quad\quad \forall i\in[n].
\end{equation}
Often one is really interested in estimating the total weight $w_I$ 
of a subset $I\subseteq [n]$ of the items, that is, $w_I=\sum_{i\in I}w_i$. 
As an estimate $\hat w_I$, we then use the sum of the
sampled items from the subset, that is, $\hat w_I=\sum_{i\in I}\hat w_I=
\sum_{i\in I\cap S} \hat w_i$. By linearity of expectation this is
also unbiased, that is, from \req{eq:single} we get
\begin{equation}\label{eq:sum}
\E[\hat w_I]=w_I \quad\quad \forall I\subseteq[n].
\end{equation}
We are particularly interested in cases where the subset $I$ is
unknown at the time the sampling decisions are made.  For example, in
an opinion poll, the subset corresponding to an opinion is only revealed
by the persons sampled for the poll.  In the context of a huge data
base, sampling is used to reduce the data so that we can later support
fast approximate aggregations over arbitrary selected subsets
\cite{GG01,OR95,JMR05}.

Applied to Internet traffic analysis, the items could be records
summarizing the flows streaming by a router. The weight of a flow 
would be the number of bytes. The stream is very high volume so we can only
store samples of it efficiently. A subset of interest
could be flow
records of a newly discovered worm attack whose signature has just
been determined.  The sample is used to estimate the
size of the attack even though the worm was unknown at the time the
samples were chosen.  This particular example is discussed 
in \cite{JMR05}, which also shows how the subset sum sampling can be 
integrated in a data base style infrastructure for a streaming context.
In \cite{JMR05} they use the threshold sampling from \cite{DLT05} which 
is one the sampling schemes that we will analyze below.

Generally there are two things we want to minimize: (a) the number of
samples viewed as a resource, and (b) the variance as a measure
for uncertainty in the estimates.

For several sampling schemes, we already understand the optimality
with respect to the sum of the individual variances
\begin{equation}\label{eq:var-sum}
\SV=\sum_{i\in[n]} \Var[\hat w_i]
\end{equation}
as well as the variance of the total sum
\begin{equation}\label{eq:sum-var}
\VS=\Var[\hat w_{[n]}]\quad \left(=\Var[\sum_{i\in[n]} \hat w_i]\right)
\end{equation}
However, what we are really interested in is the estimation of subsets of
arbitrary sizes.

Before continuing, we note that there is an alternative use of sampling
for subset sum estimation in data bases; namely where data
are organized to generate a sample from any selected subset. Generating
such samples on-the-fly has been studied with different sampling
schemes in \cite{ADLT05,Coh99,HHW97}. When each subset gets its own
sample, we are only interested in the variance of totals $\VS$.

In this paper, we generate the sample first, and then we use this
sample to estimate the weight of arbitrary subsets. As discussed in
\cite{JMR05}, this is how we have to do in a high volume streaming
context where items arrive faster and in larger quantities than can be
saved; hence where only a sample can be stored efficiently. The sampling
first is also relevant if we want to create a reduced approximate version of a
large data ware house that can be downloaded on smaller device.

\subsection{Performance measure}
The purpose of our sampling is later to be able to estimate arbitrary subset
sums.  With no advance knowledge of the subsets of interest, a natural
performance measure is the expected variance for a random subset. We
consider two distributions on subsets:
\begin{description}
\item[$\Sm m$] denoting the uniform distribution on subsets
of size $m$.
\item[$\Sp p$] denoting the distribution on subsets where each
item is included independently with probability $p$.
\end{description}
Often we are interested in smaller subsets with $p=o(1)$ or $m=o(n)$.
The corresponding expected variances are denoted
\begin{eqnarray*}
{\Vm m} &=& \E_{I\leftarrow \Sm m}[\Var[\hat w_I]]\\
{\Wp p}&=& \E_{I\leftarrow \Sp p}[\Var[\hat w_I]]
\end{eqnarray*}
Note that  $\Vm 1=\SV/n$ and  $\Vm n=\Wp 1=\VS$. 

We are not aware of any previous analysis of the average 
variance of subset sum estimation.

\subsection{A basic theorem}\label{S:theorem}
Our basic theorem below states that our subset sum variances are simple 
combinations of
$\SV$ and $\VS$. The quantities $\SV$ and $\SV$ are often quite
easy to analyze, and from them we immediately derive any ${\Vm m}$.
\begin{theorem}\label{thm:main} For any sampling scheme, we have
\begin{eqnarray}
\Vm m&=&\frac mn\left(\frac{n-m}{n-1}\,\SV+\frac{m-1}{n-1}\,\VS\right)\label{eq:main-mn}\\
\Wp {p}&=&p\left((1-p)\SV+p\VS\right).\label{eq:main-p}
\end{eqnarray}
\end{theorem}
Theorem \ref{thm:main} holds for arbitrarily correlated random
estimators $\hat w_i, i\in[n]$ with $E[\hat w_i]=w_i$. That is, we
have an arbitrary probability space $\Phi$ over functions $\hat w$
mapping indices $i\in [n]$ into estimates $\hat w_i$.  Expectations
and variances are all measured with respect to $\Phi$.  The only
condition for our theorem to hold true is that the estimate of a
subset is obtained by summing the estimates of its element, that is,
$\hat w_I=\sum_{i\in I}\hat w_i$.

One nice consequence of \req{eq:main-mn} is that 
\[{\Vm m}\geq m\frac{n-m}{n-1}\,\Vm 1\]
This means that no matter how much negative covariance we
have, on the average, it reduces the variance by at most a factor
$\frac{n-1}{n-m}$.

A nice application of \req{eq:main-p} is in connection with a random
partition into $q$ subsets where each item independently is assigned a 
random subset. A given subset includes each item with
probability $p=1/q$, so by linearity of expectation, the expected
total variance over all sets in the partition is
\[q\cdot {\Wp p}=\left((1-p)\SV+p\VS\right)\]

\subsection{Known sampling schemes}
We will apply Theorem~\ref{thm:main} to study the optimality of some
known sampling schemes with respect to the average variance of subset sum
estimation.  Below we first list the schemes and discuss.
what is known about $\SV$ and $\VS$. Our
findings with Theorem \ref{thm:main} will be summarized in the next
subsection.

Most of the known sampling schemes use Horvitz-Thompson estimators: if
item $i$ was sampled with probability $p_i$, it is assigned an
estimate of $\hat w_i=w_i/p_i$. Horvitz-Thompson estimators are
trivially unbiased.

For now we assume that the weight $w_i$ is known before the 
sampling decission is made. This is typically not the case in survey
sampling. We shall return to this point in Section \ref{sec:survey}.

\paragraph*{Uniform sampling without replacement (U$+$R)}
In uniform sampling without replacement, 
we pick a sample of $k$ items
uniformly at random. If item $i$ is sampled it gets 
weight estimate $\hat w_i=w_i n/k$. We denote this scheme $\Uwor k$.

\paragraph*{Probability proportional to size sampling with replacement (P$+$R)}
In probability proportional to size sampling with replacement, each
sample $S_j\in [n]$, $j\in[k]$, is independent, and equal to $i$ with
probability $w_i/w_{[n]}$. We say
that $i$ is sampled if $i=S_j$ for some $j\in[k]$. This happens with
probability $p_i=1-(1-w_i/w_{[n]})^k$. If $i$ is now sampled,
we use the Horvitz-Thompson estimator $\hat w_i=1/p_i$. 
We denote this scheme $\Pwr k$.

\paragraph*{Threshold sampling (THR)}
The threshold sampling is a kind of Poisson
sampling. In Poisson sampling, each item $i$ is picked independently
for $S$ with some probability $p_i$. For unbiased estimation, we use
the Horvitz-Thompson estimate $\hat w_i=w_i/p_i$ when $i$ is
picked. 

In threshold sampling 
we pick a fixed threshold $\tau$. For the sample
$S$, we include all items with weight bigger than $\tau$. Moreover, we include
all smaller items with probability $w_i/\tau$. 
Sampled items $i\in S$
have the Horvitz-Thompson estimate 
$\hat w_i=w_i/p_i=w_i/\min\{1,w_i/\tau\}=\max\{w_i,\tau\}$. 
With $k=\sum_i\min\{1,w_i/\tau\}$ the expected  number of samples, we denote
this scheme $\THR k$. Threshold sampling is known to minimize $\SV$ relative
to the expected number of samples.

In survey sampling, one often makes the simplifying assumption that if
we want $k$ samples, no single weight has more than a fraction $1/k$
of the total weight \cite[p. 89]{SSW92}.  In that case threshold
sampling is simply Poisson sampling with probability proportional to
size as described in \cite[p. 85--87]{SSW92}. More precisely, the
threshold becomes $\tau=w_{[n]}/k$, and each item is sampled with
probability $w_i/\tau$. We are, however, interested in the common case
of heavy tailed distributions where a one or a few weights dominate
the total \cite{AFT98,PKC96}. The name ``threshold sampling'' for the
general case parameterized by a threshold $\tau$ is taken from
\cite{DLT05}.

\paragraph*{Systematic threshold sampling (SYS)}
We consider the general version of systematic sampling where
each item $i$ has an individual sampling probability $p_i$, and
if picked, a weight estimate $w_i/p_i$. Contrasting Poisson sampling,
the sampling decisions are not independent. Instead we pick a
single uniformly random number $x\in[0,1]$, and include $i$ in $S$ if
and only if for some integer $j$, we have 
\[\sum_{h<i}p_i \leq j+x<\sum_{h\leq i}p_i\]
It is not hard to see that $\Pr[i\in S]=p_i$.
Let $k=\sum_{i\in[n]}p_i$ be the expected number of samples. Then the
actual number of samples is either $\lfloor k\rfloor$ or $\lceil
k\rceil$. In particular, this number is fixed if $k$ is an integer.
Below we assume that $k$ is integer.

In systematic threshold sampling we perform systematic sampling
with exactly the same sampling 
probabilities as in threshold sampling, and denote this scheme
$\SYS k$. Hence for each item $i$, we have identical 
marginal distributions $\hat w_i$ with $\THR k$ and $\SYS k$.

\paragraph*{Priority sampling (PRI)}
In priority sampling from \cite{DLT04c} 
we sample a specified number of $k<n$ samples.
For each item, a we generate a uniformly random number $r_i\in(0,1)$,
and assign it a priority $q_i=w_i/r_i$. We assume these priorities are
all distinct. The $k$ highest priority items are sampled. We call the
$(k+1)$th highest priority the threshold $\tau$. Then $i$ is sampled
if and only if $q_i>\tau$, and then the weight estimate is $\hat
w_i=\max\{\tau,w_i\}$. This scheme is denoted $\PRI k$.

Note that the weight estimate $\hat
w_i=\max\{\tau,w_i\}$ depends on the random variable $\tau$ which is
defined in terms of all the priorities. This is not a Horvitz-Thompson
estimator. In \cite{DLT04c} it is proved that this estimator is unbiased, and
that there is no covariance between individual
estimates for $k >1$.

\subsection{Variance optimality of known sampling schemes}\label{SS:optimality}
Below we compare $\Vm m$ and $\Wp p$ for the different sampling
schemes. Using Theorem \ref{thm:main} most results are
derived quite easily from existing knowledge on $\SV$ and $\VS$.
The derivation including the relevant existing knowledge will be 
presented in Sections \ref{S:near-optimal}--\ref{S:anti-optimal}.

When comparing different sampling schemes, we use
a superscript to specify which sampling scheme is used. For example
$\Vm m^\Phi< \Vm m^\Psi$ means that the sampling scheme
$\Phi$ obtains a smaller value of $\Vm m$ than does $\Psi$.

For a given set of input weights $w_1,...w_n$, we think abstractly of
a sampling scheme as a
probability distribution $\Phi$ over functions $\hat w$ mapping items
$i$ into estimates $\hat w_i$. We require unbiasedness in the sense
that $\E_{\hat w\leftarrow\Phi}[\hat w_i]=w_i$.  For a given
$\hat w_i\in\Phi$, the number of samples is the number of
non-zeroes. For any measure over sampling schemes, we use a 
superscript $\OPT k$
to indicate the optimal value over all sampling schemes using an expected
number of at most $k$ samples. For example, $\Vm m^{\OPT k}$ is the minimal
value of $\Vm m^\Phi$ for sampling schemes $\Phi$ using an expected
number of at most $k$ samples.

\paragraph*{Optimality of SYS, THR, and PRI}
For any subset size $m$ and sample size $k$, we get
\begin{equation}\label{eq:opt-sys-thr-mn}
\Vm m^{\OPT k}=\Vm m^{\SYS k}=\frac{n-m}{n-1}\,\Vm m^{\THR k}
\end{equation}
The input weights $w_1,...,,w_n$ 
where arbitrary, so we conclude that systematic threshold sampling optimizes
${\Vm m}$ for any possible input, subset size $m$, and sample size, against
any possible sampling scheme. For contrast, threshold sampling is always 
off by exactly a factor $\frac{n-1}{n-m}$.

Similarly, for any subset inclusion probability $p$, we get that
\begin{equation}\label{eq:opt-sys-thr-p}
{\Wp p}^{\OPT k}={\Wp p}^{\SYS k}=(1-p)\,{\Wp p}^{\THR k}
\end{equation}

From \cite{Sze06}, we get that 
\begin{eqnarray}\label{eq:pri-thr-mn}
\Vm m^{\PRI {k+1}}&\leq &\Vm m^{\THR {k}}\leq \Vm m^{\PRI k}\\
\label{eq:pri-thr-p}
W_p^{\PRI {k+1}}&\leq& W_p^{\THR {k}}\leq W_p^{\PRI k}
\end{eqnarray}
Hence, modulo an extra sample, priority sampling is as good as
threshold sampling, and hence at most a factor $\frac{n-1}{n-m}$ or
$1/(1-p)$ worse than the optimal systematic threshold sampling.

\paragraph*{Anti-optimality of U$-$R and P$+$R} We argue 
that standard sampling schemes such as uniform sampling and
probability proportional to size sampling with replacements may be
arbitrarily bad compared with the above sampling schemes. The main
problem is in connection with heavy tailed weight distributions where
we likely have one or a few dominant weights containing most of the
total weight.  With uniform sampling, we are likely to miss the
dominant weights, and with probability proportional to size sampling
with replacement, our sample gets dominated by copies of the dominant
weights. Dominant weights are expected in the common case of heavy
tailed weight distributions \cite{AFT98,PKC96}.

We will analyze a concrete example showing that these classic
schemes can be arbitrarily bad compared with the above near-optimal
schemes. The input has a large weight $w_n=\ell$
and $n-1$ unit weights $w_i=1$, $i\in [n-1]$. We are
aiming at $k$ samples. We assume that
$\ell\gg n\gg k\gg 1$ and $\ell\geq k^2$. Here $x\gg y\iff x=\omega(y)$.
For this concrete example, we will show that
\begin{eqnarray*}
\Vm m^{\OPT k}&\approx& (n-m)m/k\\
V^{\Uwor k}_m&\gappr& \ell^2m/k\\
\Vm m^{\Pwr k}&\gappr & \ell m/k
\end{eqnarray*}
Here $x\approx y \iff x=(1\pm o(1)) y$ and $x\gappr y \iff x\geq (1-o(1)) 
y$. A corresponding set of relations can
be found in terms of $p$, replacing $n-m$ with $n(1-p)$ and $m$ with
$pn$.
We conclude that uniform sampling with replacement is a factor $\ell^2/n$
from optimality while probability proportional to size sampling with replacement is a factor
$\ell/n$ from optimality. Since $\ell\gg n$ it follows that both schemes
can be arbitrarily far from optimal. 

\subsection{Discussion} 
One of our conclusions above is that
systematic threshold sampling is optimal for the average subset
variances no matter the subset size $m$ or inclusion probability $p$.
However, there may be scenarios where some sampling schemes are 
not appropriate. In the Section \ref{sec:appliation} we will study a streaming
scenario ruling out both threshold and systematic threshold sampling, leaving
us with priority sampling among the near-optimal schemes. From
\req{eq:opt-sys-thr-p} and \req{eq:pri-thr-p}, we get that
\begin{equation}\label{eq:pri-opt}
\Vm m^{\PRI {k+1}}\leq \frac{n-1}{n-m}\,\Vm m^{\OPT k}
\end{equation}
Even if we don't know what the optimal appropriate scheme is, this
inequality provides a limit to the improvement with any possible
scheme. In particular, if $k$ is not too small, and $m$ is not too
close to $n$, there is only limited scope for improvement.

\subsection{Contents} The rest of the paper is divided
as follows: In Section \ref{sec:appliation} we 
discuss our results in concrete application scenarios including
survey sampling and related experimental work. In Section \ref{S:proof}
we prove
Theorem \ref{thm:main}. In Sections 
\ref{S:near-optimal}--\ref{S:anti-optimal} we will
derive the optimability results. In Section 
\ref{sec:bias} we discuss extensions to biased sampling, and
finally we have some concluding remarks in Section \ref{sec:conclussion}.

\section{Application scenario and related experimental work}
\label{sec:appliation}
In this section, we shall discuss an important Internet related
application where systematic and threshold sampling are less
appropriate, hence where the better choice is to settle for
near-optimality of priority sampling.

The setup of the scenario in this section is taken from \cite{DLT04j,JMR05}. It
serves to contextualize the preceding optimality results in a realistic
context.  We will also mention related experimental work from
\cite{DLT04j} complementing the analytic results of this paper.

\subsection{Reservoir sampling}
We are here focusing on reservoir sampling (c.f. \cite{FMR62} and
\cite[p. 138--140]{Knu69})
for a stream of weighted items. In reservoir sampling, the items
arrive one by one, and a reservoir maintains a sample $S$ of the items
seen thus far. When a new items arrives, it may be included in the
sample $S$ and old samples may be dropped from $S$. Old items
outside $S$ are not reconsidered. Reservoir sampling addresses two
issues:
\begin{itemize}
\item The streaming issue \cite{Mut05} where we 
want to compute a sample from a huge stream that passes by only
once, when the memory available to us is limited. 
\item The incremental data structure issue of maintaining a sample as new items
are added. In our case, we use the sample to provide quick
estimates of sums over arbitrary subsets of the items seen thus far.
\end{itemize}
The reader is referred to \cite{JMR05} for a description of how
reservoir sampling and subset sum estimation can be integrated in
a data base style infrastructure for a streaming context.

\subsection{Relation to survey sampling}\label{sec:survey}
The above set-up is similar to that of classic survey sampling 
(see, e.g. \cite{SSW92}). However, in survey sampling, typically, we do not
know the weight $w_i$ of an item $i$ unless we sample it. Instead we have 
free access to an auxiliary variable $u_i$ that is correlated with $w_i$, and
use $u_i$ to determine the sampling probability $p_i$ for item $i$.
For example, if the item $i$ is a house and $w_i$ is house hold income,
then $u_i$ could be an approximation of $w_i$ based on the address.
We can then use $u_i$ to determine the sampling probability $p_i$ for item $i$.
The weights $w_i$ will only be found for the items sampled.

The previously discussed techniques provide an estimate $\hat u_i$
of the known variable $u_i$, and then we use $\hat w_i=w_i\hat u_i/u_i$
as an estimator for $w_i$. If $\hat u_i$ is an unbiased estimator, 
then so is $\hat w_i$, that is, $\E[\hat w_i]=w_i\E[\hat u_i]/u_i=w_i$. Also, if $\hat u_i$ is a Horvitz-Thompson 
estimator, then so is $\hat w_i$, that is, if $i$ is sampled, 
then $\hat w_i=w_i\hat u_i/u_i=w_i(u_i/p_i)u_i= w_i/p_i$.

In survey sampling, the main challenge is often to estimate the total
$w_{[n]}$ based on the sampled weights. They often have an analysis of
$\VS$ assuming that $w_i=u_i$, and then they use this to indicate that
a scheme will be good if $w_i \approx u_i$. For example, it is
known that  $\VS^{\SYS k}=0$  \cite[pp. 88,96,97]{SSW92}, and that
threshold sampling minimizes $\VS$ among all Poisson sampling
schemes   \cite[p. 86]{SSW92}.

Knowing the exact weight comes in naturally in computer
science when the purpose of the sampling is to reduce a
large data set so that we can later support
fast approximate aggregations over arbitrary subsets.
For example, this idea is used in data bases \cite{CMN99,GG01,OR95,JMR05}. 
Nevertheless, there could be cases where sampling is made with
one weight $u_i$ in mind, but later used for another weight $w_i$.
This case is treated as in survey sampling. In case of heavy tailed
distributions, uniform sampling is basically useless. Hence it is
very important that $u_i$ is large when $w_i$ is large.

Our context is that of a large stream of weighted items passing by. When
item $i$ passes by, we get to see its weight $w_i$. If our goal
was to compute $w_{[n]}$, we would simply accumulate the weights in a counter.
Hence, in our context, the challenge of survey sampling is trivial. 

One thing that makes reservoir sampling hard is that sampling decisions
are made on-line. This rules out off-line sampling
schemes such as Sunter's method \cite{Sun77},\cite[p. 93--97]{SSW92} 
where we have to sort 
all the items before any sampling decisions are made.

A cultural difference between survey sampling and our case is
that survey sampling appears less focused on heavy tailed distribution.
For threshold or systematic threshold sampling one can then assume that the threshold
is bigger than the maximal weight, hence that these schemes use
probabilities proportional to size. In our kind of applications, heavy
tailed distributions are very prominent \cite{AFT98,PKC96}.

\subsection{Internet traffic analysis}\label{S:examples}
With a concrete Internet example, we will
now illustrate the selection of subsets
and the use of reservoir sampling for estimating the sum over
these subsets. For the selection, the basic point is that an item, 
besides the weight,
has other associated information, and selection of an item may
be based on all its associated information. As stated in \req{eq:sum},
to estimate the total
weight of all selected items, we sum the weight estimates of
all selected sampled items. 

Internet routers export information about transmissions of
data passing through. These transmissions are called flows.  A flow
could be an ftp transfer of a file, an email, or some other collection
of related data moving together. A flow record is exported with
statistics such as application type,
source and destination IP addresses, and the number of packets and
total bytes in the flow.  We think of the byte size as the weight.

We want to sample flow records in such a way that we can answer
questions like how many bytes of traffic came from a given customer or
how much traffic was generated by a certain application. Both of these
questions ask what is the total weight of a certain selection
of flows.  If  we knew in advance of
measurement which selections were of interest, we could have a counter
for each selection and increment these as flows passed by.
The challenge here is that we must not be constrained to selections
known in advance of the measurements. This would preclude
exploratory studies, and would not allow a change in routine questions
to be applied retroactively to the measurements. 

A killer example where the selection is not known in advance was the
tracing of the {\em Internet Slammer worm} \cite{Slammer}. It turned
out to have a simple signature in the flow record; namely as being udp
traffic to port 1434 with a packet size of 404 bytes.  Once this
signature was identified, the worm could be studied by selecting
records of flows matching this signature from the sampled
flow records. 

We note that data streaming algorithms have been developed that
generalize counters to provide answers to a range of selections such
as, for example, range queries in a few dimensions
\cite{Mut05}. However, each such method is still restricted to a
limited type of selection to be decided in advance of the measurements.

\subsection{Experiments on real Internet data}
\begin{figure}
\begin{center}
\epsfig{file=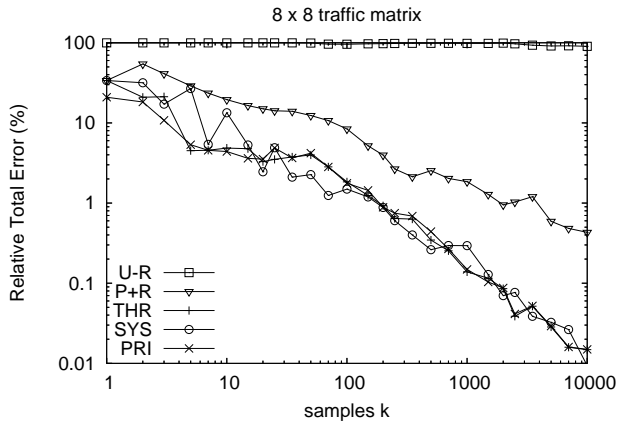,width=3.25in}
\end{center}
\caption{Estimation of $8\times8$ traffic matrix.}\label{fig:matrix}
\end{figure}

In \cite{DLT04j}, the above Internet application is explored with
experiments on a stream segment of 85,680  flow records exported
from a real Internet gateway router. These items were heavy tailed with a
single record representing 80\% of the total weight. 
Subsets considered were entries of an $8\times 8$ traffic matrix, as well as a
partition of traffic into traffic classes such as ftp and dns
traffic. Figure \ref{fig:matrix} shows the results for the
$8\times 8$ traffic matrix with all the above mentioned
sampling schemes (systematic threshold sampling was not included in \cite{DLT04j},
but is added here for completeness). The figure shows the 
relative error measured as the sum of 
errors over all $64$ entries divided by the total traffic.
The error is a function of the number $k$ of samples, except
with THR, where $k$ represents the expected number of samples.

We note that U$-$R is worst. It has an error close to 100\% because
it failed to sample the large dominant item. The P$+$R is much better
than U$+$R, yet much
worse than the near-optimal schemes PRI, THR, and SYS. 
To qualify the difference, note that P$+$R use about
50 times more samples to get safely below a 1\% relative error.

Among the near-optimal schemes, there is no clear winner. From our
theory, we would not expect much of a difference.  We would expect THR
to be ahead of PRI by at most one sample.  Also, we are studying a
partitioning into $64$ sets, and then, as noted in Section
\ref{S:theorem}, the average variance advantage of SYS is a factor
$1-1/64$, which is hardly measurable.

The experiments in Figure \ref{fig:matrix} are thus consistent with our
theory. The strength of our mathematical results is that we now know
that no one can ever turn up with a different input or a new sampling
scheme and perform better on the
average variance.  Conversely, experiments with real
data could illustrate subsets with relevant special properties that are
far from the average behavior.

\subsection{Resource constrained reservoir sampling}\label{S:k-sampling}
Our analysis names systematic threshold sampling the best possible sampling
scheme. However, in reservoir sampling we often have a resource
bound on the number $k$ of samples we can store, e.g., we may only
have a certain amount of memory available for the samples.
Priority sampling
is ideally suited for this context in that a standard priority queue
can maintain the $k+1$ items of highest priority (when a new item
arrives, it is first assigned a priority, then it is added to the
queue, and finally we remove the item of smallest priority item from
the queue in $O(\log k)$ time. 

However, with both threshold sampling and priority sampling it appears that
we need to know the threshold $\tau$ in advance (item $i$ is sampled with
probability $\min\{1,w_i/\tau\}$). This threshold $\tau$ is a function
of all items such that \linebreak[4]$\sum \min\{1,w_i/\tau\}=k$. Hence $\tau$ can
only be determined after the whole stream has been investigated.

As described in \cite{DLT04j} it is possible, though a bit more complicated,
to adapt threshold sampling for a stream to provide an expected number of $k$
samples. The essential idea is that we increase the threshold as we move
along the stream in such away that it always gives an expected number of $k$ samples from the items seen thus far. Thus an item $i$ gets dropped from the
sample when the threshold falls below its priority.
However, if we want to be sure to no more than $k$ samples are made,
we have to shoot for substantially less than $k$ samples. For example,
to stay below $k$ with 99\% probability, using Normal approximation for larger
$k$, we should only go for an expected number of $k-2.3\sqrt k$ samples.
In contrast, 
with priority sampling, we do better than threshold sampling
with an expected number of $k-1$ samples. Thus priority sampling works
better when we are allowed at most $k$ samples.

For systematic threshold sampling, the problem is more severe because
if one changes the threshold marginally, it may completely change
the set of samples.  One could conceivably resolve this if we only
increased the threshold by an exact doubling starting. However, a
doubling of the threshold can be shown to at least double the
variance. Another objection to systematic threshold sampling in a
streaming context is that we may have a very strong correlations
between items in a subset depending on how they are placed in the
stream. Normally, it is recommended that the items are appropriately
shuffled \cite[p. 92]{SSW92}, but that is not possible in reservoir
sampling from a stream. With threshold and priority sampling there is no such
dependence as there is no covariance between different item estimates.
As demonstrated in \cite{Tho06}, it is possible to get
good confidence bounds with priority sampling and threshold sampling
so that we statistically know when we get good estimates for a
subset. The correlation between items in systematic threshold sampling
prevents us from providing good confidence intervals, so even if
systematic threshold sampling gives better variance on the average, we
have no way of knowing if we get these good estimates for a concrete
subset.

Thus, among our near optimal sampling schemes, priority sampling is
the most appropriate for resource constrained reservoir sampling.

\subsection{On the suboptimality of priority sampling}
Recall from \req{eq:pri-opt} that
\[\Vm m^{\PRI {k+1}}\leq \frac{n-1}{n-m}\,\Vm m^{\OPT k}\]
In our Internet application we typically have
thousands of samples. Hence we are not concerned about the difference
between $k$ and $k+1$ samples.

The factor $\frac{n-1}{n-m}$ is only significant for larger sets
$m$. However, for larger sets, we expect to do great anyway because
they relatively speaking have much smaller errors. More precisely,
we typically expect that we have plenty of samples go get very a good
estimate of the total, or in other words, that the relative
standard deviation $\eps_{n:n}=\sqrt{\VS}/w_{[n]}$
for the total is very small.

Since priority
sampling has no covariance, $\Vm m=m/n\cdot \VS$. At the same time, the 
average subset sum is $m/n\cdot w_{[n]}$. For a subset achieving both
of these averages, the relative standard deviation would be
\[\eps_{m:n}=\frac{\sqrt{m/n\cdot \VS}}{m/n\cdot w_{[n]}}=\sqrt{n/m}\,\eps_{n:n}\]
However, if $\sqrt{n/m}$ is big, then the optimality factor
$\frac{n-1}{n-m}$ is close to $1$.

Thus, it is when our variance is expected comparatively small that our
relative distance to OPT is greatest, the most extreme being in the
estimation of the total. The estimate of the total has the smallest
relative standard deviation, but since it is positive, it is
infinitely worse than $\Vm n^{\OPT k}=0$.

Another case where we do not need to worry so much about the
non-optimality factor $\frac{n-1}{n-m}$ is if we are interested in the
relative weight of a subset $I$ of size $m$. As an estimator, we use
$\hat w_I/\hat w_{[n]}$. If $m> n/2$, we note that $\hat w_I/\hat
w_{[n]}=1-\hat w_{[n]\setminus I}/\hat w_{[n]}$.  Most of the error in
this estimate stems from the estimate $\hat w_{[n]\setminus I}$ of the
small set $[n]\setminus I$, but for this small set size, we are at
most a factor $\frac{n-1}{n-(n/2+1)}<2$ from optimality.

As discussed in Section \ref{S:k-sampling}, we do not know if
there is a scheme performing better than priority sampling in practice
in the context of resource constrained reservoir sampling. The conclusion
of this section is that even if there is a better scheme, it is not
going to help us much.

\section{Proof of basic theorem}\label{S:proof}
In this section we prove \req{eq:main-mn}
\[{\Vm m}=\frac mn \left(\frac{n-m}{n-1}\,\SV+\frac{m-1}{n-1}\,\VS\right)\]
and \req{eq:main-p}
\[\Wp p=p\left((1-p)\SV+p\VS\right).\]
By the definitions of variance and covariance, for any subset $I\subseteq[n]$,
\[\Var[\hat w_I]=A_I+B_I\]
where
\begin{eqnarray*}
A_I&=&\sum_{i in I} \Var[\hat w_i]\\
B_I&=&\sum_{i,j \in I, i\neq j} \CoV[\hat w_i,\hat w_j]
\end{eqnarray*}
Suppose $I$ is chosen uniformly at random among subsets of
$[n]$ with $m$ element.
Then for any $i$, $\Pr[i \in I]=m/n$, so by linearity of expectation, 
\begin{eqnarray*}
\E[A_I]&=&\sum_{i \in [n]}\Pr[i\in I]\Var[\hat w_i]\\
&=&m/n\cdot A_{[n]}.
\end{eqnarray*}
Also, for any $j\neq i$, $\Pr[i,j\in I]=m/n\cdot(m-1)/(n-1)$, 
so by linearity of 
expectation, 
\begin{eqnarray*}
\E[B_I]&=&\sum_{i,j \in [n], i\neq j}\Pr[i,j\in I]\CoV[\hat w_i,\hat w_j]\\
&=&m/n\cdot (m-1)/(n-1)\cdot B_{[n]}.
\end{eqnarray*}
Thus 
\begin{equation}\label{eq:*}
\E[\Var[\hat w_I]] = m/n\cdot A_{[n]} + m/n\cdot (m-1)/(n-1)\cdot B_{[n]}
\end{equation}
By definition, $A_{[n]}=\SV$. Moreover, by \req{eq:*},
\[\VS=A_{[n]} + B_{[n]}\]
so 
\[B_{[n]} = \VS-\SV.\]
Consequently,
\begin{eqnarray*}
{\Vm m} &=&\E[\Var[\hat w_I]]\\
&=& \frac mn\,\SV + \frac mn\frac{m-1}{n-1}(\VS-\SV)\\
&=& \frac mn\,\left(\frac{n-m}{n-1}\,\SV+\frac{m-1}{n-1}\,\VS\right)
\end{eqnarray*}
This completes the proof of \req{eq:main-mn}.

The proof of \req{eq:main-p} is very similar. In this case, each
$i\in[n]$ is picked independently for $I'$ with probability $p$.
By linearity of expectation, 
\[\E[A_I]=p A_{[n]}.\]
Also, for any $j\neq i$, $\Pr[i,j\in I]=p^2$, 
so by linearity of 
expectation, 
\[\E[B_I]=p^2 B_{[n]}.\]
Thus 
\begin{eqnarray*}
V'_p&=&\E[\Var[\hat w_I]]\\
&=&p A_{[n]} + p^2 B_{[n]}\\
&=&p \SV + p^2(\VS-\SV)\\
&=&p((1-p)\SV + p\VS)\\
\end{eqnarray*}
This completes the proof of \req{eq:main-p}, hence of Theorem \ref{thm:main}.

\section{Near-optimal schemes}\label{S:near-optimal}
We will now use Theorem~\ref{thm:main} to study the average 
variance (near) optimality 
of subset sum estimation with threshold sampling,
systematic threshold sampling, and priority sampling for any possible set of input weights.  The results are all derived based on existing knowledge
on $\SV$ and $\VS$. Below we will focus on $\Vm m$ based on random
subsets of a given size $m$. The calculations are very similar for
$\Wp p$ based on the inclusion probability $p$.

It is well-known from survey sampling that \cite[pp. 88,96,97]{SSW92} 
that systematic sampling always provides an exact
estimate of the total so $\VS^{\SYS k}=0$. Since
variances cannot be negative, we have
\[\VS^{\SYS k}=0=\VS^{\OPT k}.\]
It is also known from survey sampling \cite[p. 86]{SSW92} 
that threshold sampling minimize $\VS$ among
all Poisson sampling schemes. In \cite{DLT04j} it is further argued
that threshold sampling minimizes $\SV$ over all possible
sampling schemes, that is, $\SV^{\THR k}=\SV^{\OPT k}$. Since
systematic threshold sampling uses the same marginal distribution for the 
items, we have
\[\SV^{\THR k}=\SV^{\SYS k}=\SV^{\OPT k}.\]
Since $\SYS k$ optimizes both $\SV$ and $\VS$ we conclude \req{eq:main-mn}
that it optimizes $\Vm m$ for any subset size
$m$. More precisely, using
\req{eq:main-mn}, we get
\begin{eqnarray*}
\Vm m^{\SYS k}&=&\frac mn \left(\frac{n-m}{n-1}\,V^{\SYS k}_1+\frac{m-1}{n-1}\,V^{\SYS k}_n\right)\\
&=&\frac mn \left(\frac{n-m}{n-1}\,\SV^{\OPT k}+\frac{m-1}{n-1}\cdot 0\right)\\
&\leq &\Vm m^{\OPT{k}} \leq \Vm m^{\SYS k}.
\end{eqnarray*}
Hence
\[\Vm m^{\SYS k}=\Vm m^{\OPT k}.\]
As mentioned above we have $\SV^{\THR k}=\SV^{\SYS k}$. Moreover,
threshold sampling has no covariance between individual estimates,
so 
\[\Vm m^{\THR k}=\frac mn\,\SV^{\THR k}=\frac mn\,\SV^{\OPT k}.\]
But in the previous  calculation, we saw that 
\[\Vm m^{\SYS k}=\frac mn \frac{n-m}{n-1}\,\SV^{\OPT k}\]
Hence we conclude that 
\[\Vm m^{\SYS k}=\frac{n-m}{n-1}\,\Vm m^{\THR k}\]
This completes the proof of \req{eq:main-mn}. A very similar
calculation establishes \req{eq:main-p}.

In \cite{Sze06} it is proved that
\[\SV^{\PRI {k+1}}\leq \SV^{\THR {k}}\leq\SV^{\PRI k}\]
Moreover, for any scheme $\Phi$ without covariance, we have 
\[\Vm m^{\Phi}=\frac mn\,\SV^{\Phi}\]
Since both threshold and priority sampling have no covariance,
we conclude \req{eq:pri-thr-mn}
\[\Vm m^{\PRI {k+1}}\leq \Vm m^{\THR {k}}\leq\Vm m^{\PRI k}\]
The proof of \req{eq:pri-thr-p} is similar based on 
$\Wp p^{\Phi}=p\,\SV^{\Phi}$.

\section{Anti-optimal schemes}\label{S:anti-optimal}
Below we will analyze a concrete example showing that the classic
schemes of uniform sampling without replacement and probability proportional to size sampling
with replacement can be arbitrarily bad compared with the above near-optimal
schemes. 

The concrete example consists of $n-1$ unit weights $w_i=1$, $i\in [n-1]$ and a
large weight $w_n=\ell$. We are aiming at $k$ samples. We assume that
$\ell\gg n\gg k\gg 1$ and that $\ell\gg k^2$.

As in the last section, we focus on the subset size $m$ rather than
the inclusion probability $p$.

\subsection{Threshold sampling}
We will now analyze the variance with threshold sampling for
the bad example. The variance with systematic and priority sampling
will then follow from \req{eq:opt-sys-thr-mn} and \req{eq:pri-thr-mn}.

Threshold sampling ($\THR k$) will use the threshold
$\tau=\frac{n-1}{k-1}$. This will pick the large weight $w_n=\ell$ with
probability $p_n=1$ and weight estimate $\hat w_n=w_n$.
Hence $\Var[\hat w_n]=0$. Each unit weight $w_i$, $i<n$, is then picked
with probability $p_1=\frac{k-1}{n-1}$ and estimate
$1/p_1=\frac{n-1}{k-1}$. The variance of the estimate for a unit weight item
is then $p_1(1-p_1)/p_1^2=(1-p_1)/p_1 =\frac{n-k}{k-1}$, so the total
variance is $\SV=\frac{(n-1)(n-k)}{k-1}\approx \frac{n^2}k$. 
Since there is no co-variance, we conclude for any subset size $m\leq n$ that 
\[\Vm m^{\THR k}=m/n\cdot \SV^{\THR k}\approx mn/k\]
From \req{eq:opt-sys-thr-mn} we get that 
\[\Vm m^{\SYS k}\approx \frac{n-m}{n-1}\,mn/k\approx (n-m)m/k\]
Finally, since $k=\omega(1)$ it follows from \req{eq:pri-thr-mn}
that 
\[\Vm m^{\PRI k}\in[\Vm m^{\THR k},\Vm m^{\THR {k-1}}]\approx mn/k.\]

\subsection{Uniform sampling without replacement}
In uniform sampling without replacement ($\Uwor k$), we pick a sample of
$k$ different items uniformly at random. 
As we shall see below,
the variance of uniform sampling is dominated by the variance of
estimating the large weight.

The large weight $w_n=\ell$ is picked with
probability $p_n=k/n$ and estimate $\ell/p$. 
Hence
\[\E[\hat w_n^2]=p_n(\ell/p_n)^2=n\ell^2/k.\]
It follows that 
\[\Var[\hat w_n]=\E[\hat w_n^2]-w_n^2=n\ell^2/k-\ell^2\approx n\ell^2/k\]
Hence 
\[\SV\geq \Var[\hat w_n]\approx n\ell^2/k.\]
To study the variance $\VS$ of the total sum estimate $\hat
w_{[n]}$, we note that
\[\E[\hat w_{[n]}^2]\geq
\E[\hat w_n^2]= n\ell^2/k.\]
Hence 
\[\VS=\E[\hat w_{[n]}^2]-w_{[n]}^2\geq n\ell^2/k-(\ell+n-1)^2
\approx n\ell^2/k\]
Since $\SV$ and $\VS$ are both lower bounded by 
$(1-o(1))
n\ell^2/k$, it follows
from \req{eq:main-mn} that for any subset size $m\leq n$
\[V^{\Uwor k}_m\gappr (m/n)n\ell^2/k=m\ell^2/k\]
This is roughly a factor $\ell^2$ worse than what we had with any of
the near optimal schemes.

\subsection{Probability proportional to size sampling with replacement}
In probability proportional to size sampling with replacement ($\Pwr k$), each sample $S_j\in [n]$, $j\in[k]$, is
independent, and equal to $i$ with probability $w_i/w_{[n]}$. An 
item $i$ is sampled if $i=S_j$ for some $j\in[k]$. This happens with
probability $p_i=1-(1-w_i/w_{[n]})^k$, and if $i$ is sampled,
$\hat w_i=1/p_i$. 

In our bad example, the variance with $\Pwr k$ relates to the fact
that we get mostly duplicates of the large item. The expected
number of unit samples is only $n-1/(n+\ell)$, and as a result,
we get a large variance from the unit items.

Each unit item is picked with probability 
\[p_1=1-(1-1/(\ell+n-1))^k\approx k/\ell.\]
Hence
\begin{equation}\label{eq:sv}
\SV\geq (n-1)\,\Var[\hat w_1]=(n-1)p_1(1-p_1)/p_1^2\approx 
n\ell/k.
\end{equation}
This is a factor $\ell/n$ worse than with threshold sampling.


\drop{AAAAAAAAAAAAAAAAAAAAAAAAAAAAAAA

We will now show that $\VS\gappr n\ell/k$.
To estimate the variance of $\hat w_{[n]}$ we define a
random variable $\chi$, that with high probability 
agrees with $\hat w_{[n]}$, and whose variance is easy to estimate.
Then we argue that the variance of $\hat w_{[n]}$
cannot be much smaller that of $\chi$.
For $1\le i \le k$ 
let $\chi_{i}$ be the characteristic function of the event that
$S_{i} \in [1,n-1]$. Define
\[
\chi = {\ell\over p_{n}} + 
{1\over p_1}\sum_{i=1}^{k} \chi_{i}.
\]
Notice that $\chi\neq \hat w_{[n]}$ only when $w_{n}$
is never sampled or when some item from $[1,n-1]$ is
sampled at least twice. Let $p = \Pr[\chi_i=1] = (n-1)/(\ell + n-1)\approx n/\ell$.
Since ${\ell\over p_{n}}$ is a constant, and the $\chi_{i}$s are
independent $(p, 1-p)$ Bernoulli trials,
\[
\Var [\chi] = \Var[\sum_{i=1}^{k} \chi_{i}] = k p(1-p){1\over p_1^{2}}
\approx  k {n\over \ell} {\ell^{2}\over k^{2} } =  n\ell/k,
\]
Typically two random variables that almost always 
agree may have very different variances.
But here we know more: for every fixed $1\le j \le k$
$\chi$ and $w_{[n]}$ coincide with probability 
$1-\epsilon$ for some small $\epsilon$ even when we
condition on the event that
$\sum_{i=1}^{k} \chi_{i} = j$. 
Moreover, the size of the event that $\sum_{i=1}^{k} \chi_{i} = 0$ is
small (if not, it just works to our advantage).
This implies that $\VS = \Var (\hat w_{[n]} )
\ge (1-\epsilon)\Var (\chi)\approx 
n\ell/k$.

AAAAAAAAAAAAAAAAAAAAAAAAA
}

We will now show that $\VS\gappr\SV$, or equivalently,
that
\[\SV-\VS=o(n\ell/k).\]
By definition
\[\VS=\SV+2\sum_{i>1}\left(
\E[\hat w_i\hat w_{[i-1]}]-w_iw_{[i-1]}\right),\]
so
\begin{eqnarray}
\SV-\VS&=&
2\sum_{i>1}\left(w_iw_{[i-1]}-\E[\hat w_i\hat w_{[i-1]}]\right)\nonumber\\
&=&2\sum_{i>1}\left(w_i(w_{[i-1]}-\E[\hat w_{[i-1]}|i\in S]\right)
\label{eq:sum1}
\end{eqnarray}
To bound this sum, first we consider the term with $i=n$.
\[w_{[n-1]}=\E[\hat w_{[n-1]}]=p_n\E[\hat w_{[n-1]}| n\in S]+(1-p_n)
\E[\hat w_{[n-1]}| n\not\in S]\]
so 
\[w_{[n-1]}-\E[\hat w_{[n-1]}| n\in S]\geq (1-p_n)
\E[\hat w_{[n-1]}| n\not\in S]\]
Here $(1-p_n)=\Pr[i\not\in S]=((n-1)/(\ell+n-1))^k<
(n/\ell)^k$. Moreover $\E[\hat w_{[n-1]}| n\not\in S]\leq k/p_1\approx \ell$, 
so $(1-p_n)\E[\hat w_{[n-1]}| n\not\in S]=o(n/k)$.
Hence 
\begin{equation}\label{eq:n}
w_n\left(w_{[n-1]}-\E[\hat w_{[n-1]}|i\in S]\right)=o(\ell n/k).
\end{equation}
as desired. Next we consider $i<n$. We have 
\begin{eqnarray*}
\lefteqn{w_i(w_{[i-1]}-\E[\hat w_{[i-1]}|i\in S])}\\
&=&(i-1)-(i-1)\Pr[1\in S|i\in S]/p_1)
\end{eqnarray*}
and 
\begin{eqnarray*}
\Pr[1\in S|i\in S]/p_1&\geq &(i-1)\Pr[1\in S|S_k=i]/p_1\\
&\geq &\Pr[1\in S|S_k=i]/p_1\\
&\geq &\frac{1-(1-1/(\ell+n-1))^{k-1}}{1-(1-1/(\ell+n-1))^k}\\
&\geq &\frac{1-(1-1/\ell)^{k-1}}{1-(1-1/\ell)^k}\\
&\geq &
\frac{(k-1)(1-\frac{k-1}{2\ell})}\ell \frac \ell k \\
&\geq &
1-1/k-\frac{k-1}{2\ell}\\
&=&
1-O(1/k)\\
\end{eqnarray*}
The last derivation follows because $\ell\geq k^2$. Hence
\[(i-1)-(i-1)\Pr[1\in S|i\in S]/p_1)=O(i/k)\]
so
\begin{eqnarray}
\lefteqn{\sum_{i=2}^{n-1}
\left(w_i(w_{[i-1]}-\E[\hat w_{[i-1]}|i\in S]\right)}\nonumber\\
&=&\sum_{i=2}^{n-1}O(i/k)=O(n^2/k)=o(n\ell/k)\label{eq:units}
\end{eqnarray}
Combining \req{eq:sv}, \req{eq:sum1}, \req{eq:n}, and \req{eq:units},
we conclude that $\SV-\VS=o(n\ell/k)$, hence that
\[\VS\gappr \VS\gappr n\ell/k\]
Together with \req{eq:sv} and Theorem \ref{thm:main}, it follows 
for any set size $m$, that
\[\Vm m^{\Pwr k}\gappr (m/n) n\ell/k=m\ell/k\]
This is a factor $\ell/n$ more than $\Vm m^{\THR k}$.

\section{Biased estimators}\label{sec:bias}
So far we have restricted our attention to unbiased estimators. With
biased estimators we would consider mean square error (MSE) instead
of just variance. We note that even though a biased estimator
may give a smaller MSE than an unbiased one, there are many standard
reasons to prefer unbiased estimators. For example, if we want to combine
estimates in a sum, we can use linearity of expectation to conclude
that the sum of the estimators is unbiased if each estimator is unbiased.
Also, if we add independent unbiased estimators, the variances are
just added. With biases, we cannot just add up the mean square errors.
An example where we wish to combine independent estimators is if
we have independent samples from different streams. In the Internet
application, these streams could be flow records from different routers where
would want to combine the information in a global picture \cite{DLT05:comb}.  In other
words, a biased estimator may be OK if all we consider is a single
isolated estimate. However, as soon as we start combining estimates,
the bias may come back and haunt us.

Despite these caveats of biased estimators, we discuss them briefly
below to see how they fit into subset sum estimation.  As a concrete
example of biased estimation, suppose a sampling scheme does not
provide exact estimation of the total, but that the total is
known. Then, for each item $i$, we could use the adjusted estimator
$\hat x^{adj}_i=\hat x_i (x_{[n]}/\hat x_{[n]})$. Then the total is right
in the sense that  $\hat x^{adj}_{[n]}=x_{[n]}$. In the case of
threshold sampling with no dominant weights, this adjusted estimator
is equivalent to the
estimator suggested in \cite[p. 87]{SSW92}. The adjusted threshold sampling
estimator will bias towards large weights. However, the corresponding
adjusted uniform sampling estimator will have bias towards smaller weights.

Now, if we allow bias, how well can we do with respect to our
average mean square error? It can easily be seen that our main theorem
holds for mean square error and not just for variances. That is, with $MSE$ 
denoting mean square error instead of $V$ for variance, we get the following
generalization of \req{eq:main-mn}:
\begin{equation}\label{eq:mse}
\MSEm m=\frac mn\left(\frac{n-m}{n-1}\,\SMSE+\frac{m-1}{n-1}\,\MSES\right)
\end{equation}
In fact, our formulas generalize to any symmetric quadratic polynomial. 
As with the variance of unbiased estimators, we
can use \req{eq:mse} to compute $\MSEm m$ for a concrete sampling scheme
for which we know $\SMSE$ and $\MSES$.

Now, if we want to minimize averages and there is no requirement
of unbiasedness, the optimal performance is obtained by a concrete
sample, thus with no randomness in the sample. Assume that the
weights are in decreasing order so that $w_1$ is the largest weight.
If all we cared
about was $\MSES$, we could give some item the total weight, and
drop the rest. If all we cared
about was $\SMSE$, the optimal choice is to pick the $k$ largest
weights, using their real weights as the estimate. Then
$\SMSE=\sum_{i>k}w_i^2$. 

To optimize $\MSEm m$, we introduce a parameter $X$ for
the negative error $w_{[n]}-\hat w_{[n]}$ in the total. Then $\MSES=X^2$. To minimize
$\SMSE$, the optimal choice is to pick the $k$ largest weights,
setting the rest to $0$. For the $k$ largest weights, we distribute
the error equally, setting
$\hat w_i =w_i+(\sum_{i>k}w_i-X)/k$. 
Then $\SMSE=k((\sum_{i>k}w_i-X)/k)^2+\sum_{i>k}w_i^2=
(\sum_{i>k}w_i-X)^2/k+\sum_{i>k}w_i^2$. 
The last term is
fixed, so to optimize $\MSEm m$, we should choose $X$ so as to minimize
\[\frac{n-m}{n-1}\,(\sum_{i>k}w_i-X)^2/k+\frac{m-1}{n-1}\,X^2.\]
For $m=1$, we choose $X=\sum_{i>k}w_i$, and then $\hat w_i=w_i$ for $i\leq k$
as discussed above.

Obviously, picking the $k$ largest weights and giving them a specific
estimate is not a good ``sampling'' scheme. The above more illustrates
the danger of just looking at averages and the deceptiveness of biased
estimation. For non-random subsets such
as a large set of small items, the above scheme would always return a zero.
This kind of unfairness isn't right. Recall that we had a similar
criticism of systematic sampling in Section \ref{S:k-sampling} if we could
not shuffle the items.

An ideal sampling scheme should both have a reasonable fairness and
perform reasonably well on the average. Threshold and priority sampling
have no covariance, so all partitions have the same total variance. Here by considering partitions rather than individual
subsets, we ensure that each item is counted exactly once. Moreover,
among unbiased schemes, they essentially got within a factor
$\frac{n-m}{n-1}$ from optimality on the average variance for subsets
of size $m$, so when $m$ is not too close to $n$, this is close to ideal.

\section{Concluding remarks}\label{sec:conclussion}
As a formal measure for ability to estimate subset sums of a set
of $n$  weighted items, we suggested
for each set size $m$, to study the average variance over
all subsets:
\[{\Vm m}=\E_{I\subseteq [n], |I|=m}\left[\Var[\hat w_I]/m\right]\] 
We discovered that ${\Vm m}$ was the following simple combination of
the sum of variances $\SV$ and the variance of the total sum $\VS$:
\[{\Vm m}=\frac mn\left(\frac{n-m}{n-1}\,\SV+\frac{m-1}{n-1}\,\VS\right).\]
A corresponding formula was found for the expected variance $\Wp p$ 
for subsets including each item independently with probability $p$.

We then considered different concrete sampling schemes. The optimality of 
$\SV$ and $\VS$ was already known for some sampling schemes, and this
now allow us to derive the optimality with respect to ${\Vm m}$ for
arbitrary subset size $m$.

We found that systematic threshold sampling was optimal with respect
to ${\Vm m}$, and that threshold sampling was off exactly by a factor
$\frac{n-m}{n-1}$.  Finally, we know that priority sampling performs
like threshold sampling modulo one extra sample. We argued that this
distance to optimality is not significant in practice when we use many
samples. This was important to know in the context of resource
constrained reservoir sampling, where priority sampling is the better
choice for other reasons.

For contrast, we also showed that more classic schemes like uniform
sampling with replacement and probability proportional to size
sampling without replacement could be arbitrarily far from
optimality. The concrete example was stylistic heavy tailed
distribution.


\end{document}